# "Anomalous Solid Solution" in Ultra-High Melting Point Oxides: A New Strategy for Developing Ultra-High Temperature Thermal Protection Coatings

Yubo Wang [a,b], Hong Meng [a,*], Pengfei He [a,*], Shujun Hu [a], Chuan Sun [a], Ximing Duan [a], Xiaopeng Lu [b], Dingwang Yuan [c], Wangyu Hu [c], Xiubing Liang [a,*]

[a] *Defense Innovation Institute, Academy of Military Science, Beijing 100071, China*

[b] *School of Materials Science and Engineering, Northeastern University, Shenyang 110819, China*

[c] *College of Materials Science and Engineering, Hunan University, Changsha 410082, China*

## Abstract

The high-temperature performance of ultra-high temperature ceramics (UHTCs) in atmospheric environment is fundamentally governed by their melting points of oxidation products. Typical high-melting-point oxides, such as $ZrO_2$, undergo phase transformations at elevated temperatures, leading to structural instability. Although doping with rare-earth or transition-metal cations can suppress these transformations, it often results in a reduction in melting point, thereby limiting practical service temperature. Here, ytterbia-stabilized zirconia (YbSZ) coatings are prepared via atmospheric plasma spraying, achieving a remarkable increase in the melting point of $ZrO_2$ to approximately 2850 ℃ and raising the ultimate plasma and oxyacetylene ablation temperature up to nearly 2780 ℃ and 3200 ℃, which is the highest temperature resistance property as reported. Notably, this performance enhancement originates from a synergistic mechanism of strengthened ionic-covalent mixed bonding and improved oxygen vacancy stability. Based on these findings, the concept of "anomalous solid solution" is firstly proposed to be used in the area of ultra-high temperature protection, which provides new insights into the compositional design of UHTC systems.

[*] Corresponding authors
E-mail addresses: menghong96@163.com (Hong Meng); hepengfei93@163.com (Pengfei He); liangxb_d@163.com (Xiubing Liang)

## 1 Introduction

Thermal protection materials are indispensable components in critical aerospace systems such as hypersonic vehicles and scramjet engines[1]. Ultra-high-temperature ceramics (UHTCs) have become one of the most promising material categories in the field of thermal protection due to their outstanding high-temperature stability, high-temperature mechanical properties, and oxidation resistance[2]. Generally, UHTCs are composed of transition metal cations and anions such as oxygen, carbon, boron, and carbonitride, for example, $ZrO_2$, ZrC, $HfB_2$, and HfCN[3]. Single-component UHTCs develop a porous structure during ablation, which often fails to meet the operational requirements of extreme environments[4]. In recent years, an increasing number of studies have focused on the field of high-entropy ceramics (HECs), including oxides, carbides, borides, and silicides, and this field continues to evolve[5–7]. Building on this, some studies have proposed extending high-entropy ceramics to compositionally complex ceramics (CCCs), combining correlation and disorder in a non-stoichiometric composition space to design UHTCs with superior performance[8]. In summary, ultra-high-temperature ceramics are evolving toward higher entropy, greater complexity, and multi-component compositions.

Admittedly, through rational composition design, the performance of multi-component UHTCs has been considerably improved relative to their single-component counterparts. Extensive previous studies have demonstrated that, compared with single-component UHTCs, the maximum service temperature of multi-component UHTCs has been elevated from 1600 °C to over 2500 °C[9]. In essence, however, the ablation resistance of these ultra-high temperature ceramics under atmospheric conditions depends on the oxide scale or oxidation products formed during the reaction process. Although non-oxide systems, including borides, carbides, and carbonitrides, possess extremely high melting points[3], they inevitably oxidize in high-temperature oxidizing environments[10,11]. The maximum temperature resistance of thermal protection materials is fundamentally determined by their thermophysical properties (particularly their melting point)[12]. Therefore, studying the melting points and phase stability of oxide-based ultra-high-temperature ceramics will provide a direct approach to developing ultra-high-temperature thermal protection coatings[9].

As a representative ultra-high-temperature oxide ceramic, $ZrO_2$ has a high melting point and low thermal conductivity and exhibits excellent performance in atmospheric ablation environments. However, at high temperatures, $ZrO_2$ undergoes a phase transition from the monoclinic to the tetragonal phase, accompanied by a volume change of approximately 1.3%, which can easily lead to structural damage[9]. Extensive studies have demonstrated that doping with certain rare-earth or transition-metal cations can effectively suppress this phase transformation[13,14]. Nevertheless, such doping inevitably introduces a high concentration of oxygen vacancies, which lowers the oxide melting point and renders the reliable service temperature consistently lower than the intrinsic melting point[9]. Recent studies have shown that doping with high-valence transition-metal cations, represented by $Ta^{5+}$, can suppress the phase transformation and inhibit oxygen vacancy formation while simultaneously increasing the melting point[15]. For instance, $Ta^{5+}$ doping in $ZrO_2$ yields $Zr_{0.88}Ta_{0.12}O_2$ with a melting point as high as 2885 °C[16], but its actual ablation performance under dynamic gas flow erosion conditions has not yet been evaluated.

In addition, previous studies have indicated that lanthanide rare-earth elements, owing to their unique electronic configurations, low electronegativity, and high chemical stability, can enhance the high-temperature stability of the $ZrO_2$ system[17]. Researchers obtained the equilibrium phase diagram of the Zr-Yb-O system through thermodynamic calculations. The results showed that the solid-liquid equilibrium temperature (liquidus temperature) reaches 3091 K at a $Yb_2O_3$ content of 20.5 mol%, representing an increase over that of the pure $ZrO_2$ system[18]. We term this phenomenon—where doping with certain rare-earth elements leads to an increase rather than a decrease in the melting point—the "anomalous solid solution" concept for lanthanide elements.

In the past, research on "anomalous solid solutions" was limited to reporting the basic physical properties of oxide systems, without exploring their potential applications in ablation resistance, which could have substantially enhanced the ablation resistance of UHTCs. In this study, we pioneered the practical application of the oxide melting point elevation theory to thermal protection materials by selecting $Yb_2O_3$ stabilized $ZrO_2$ (YbSZ) as a model system and validating its effectiveness through ablation tests. Based on the composition that exhibits the highest melting point in the "anomalous solid solution", YbSZ samples with $Yb_2O_3$ molar fractions of 8%, 13%, 18%, 23%, and 28% (designated as 8YbSZ, 13YbSZ, 18YbSZ, 23YbSZ, and 28YbSZ, respectively) were prepared. The experiments successfully raised the melting point of the $ZrO_2$ system to approximately 2850 °C and elevated its ultimate ablation temperature to nearly 2780 °C. Further density functional theory (DFT) analyses reveal that this anomalous enhancement originates from the synergistic interplay between modified chemical bonding characteristics and improved oxygen vacancy stability: an appropriate amount of Yb doping introduces stronger mixed ionic-covalent bonds while simultaneously enhancing the stability of oxygen vacancy defects. Based on these findings, we propose a new strategy for the research and development of ultra-high temperature thermal protection coatings: exploiting the "anomalous solid solution" phenomenon of lanthanide rare-earth elements to create oxides with ultra-high melting points. More importantly, this concept can be extended to other ultra-high temperature ceramic systems and holds great promise for applications in ultra-high temperature thermal protection.

## 2 Results and Discussion

### 2.1 Experimental Design and Characterization Analysis of YbSZ powder and coating

To demonstrate the "anomalous solid solution" characteristics of YbSZ and to verify its ablation performance, we designed an experimental procedure comprising powder preparation, coating fabrication, and ablation testing, as illustrated in Figure 1a. First, powders for plasma spraying were prepared, and the morphology and elemental distribution of 18YbSZ were analyzed using a scanning electron microscope (SEM) and energy-dispersive X-ray spectroscopy (EDS), as shown in Figures 1b-d. The morphology and elemental distribution of the powder after spray granulation is shown in Figure 1b. This powder consists solely of interlocking nanoscale $ZrO_2$ and $Yb_2O_3$ particles, with an extremely low apparent density; specific data are provided in Table S1. Subsequently, the powder was heat-treated to facilitate the diffusion and solid solution of Zr and Yb atoms. Figure 1c shows the microstructure of the heat-treated powder. The powder surface is relatively rough, the grains have grown larger, and the sphericity has decreased. Compared with the spray-granulated powder, the distribution of Zr and Yb elements in the heat-treated powder became significantly more uniform, indicating that atomic-scale interdiffusion during this process. Although the apparent density of this powder showed a slight increase, its flowability deteriorated dramatically, making it unsuitable for plasma spraying[19]. Therefore, the powder was subjected to atmospheric plasma spheroidization to further improve its apparent density and flowability. As shown in Figure 1d, the grains of the plasma-spheroidized powder have grown larger, exhibiting good sphericity with a dense and smooth surface. After plasma spheroidization, all elements were homogeneously distributed within the powder, and no obvious elemental enrichment was observed. Figure S1 presents the particle size distribution of the plasma-spheroidized powder, which ranges from 10 to 75 μm, with a median particle size (D50) of 38.66 μm.

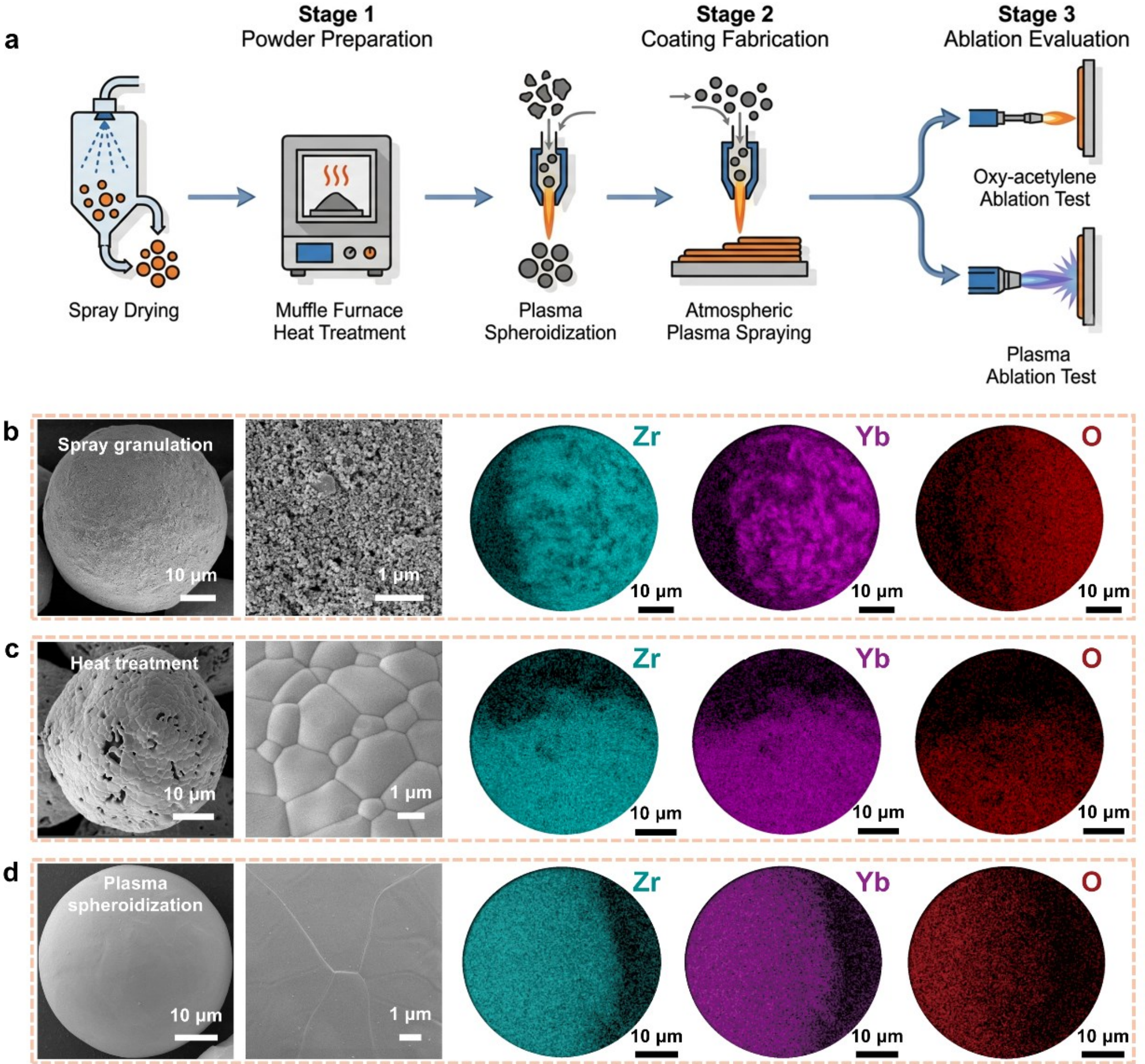

**Figure 1.** Experimental procedure and powder characterization. a) Experimental procedure: powder preparation, coating preparation, oxy-acetylene and plasma ablation tests. Morphology and elemental distribution of 18YbSZ powder prepared by b) spray granulation, c) heat treatment, and d) atmospheric plasma spheroidization.

Subsequently, YbSZ coatings fabricated using atmospheric plasma spraying were analyzed, as shown in Figure 2. Analysis of the X-ray diffraction (XRD) patterns of coatings with different $Yb_2O_3$ doping levels (Figure 2a) indicates that all samples exhibit the characteristic diffraction patterns of cubic $ZrO_2$ with the $Fm\bar{3}m$ space group. Within the detection limit, no characteristic diffraction peaks attributable to $Yb_2O_3$ or other rare earth zirconates were observed. This result clearly confirms that, within the current range of doping concentrations, the two components form a single-phase solid solution rather than a two-phase mechanical mixture. Further detailed observations revealed that, as the $Yb_2O_3$ doping concentration increased, the strongest diffraction peak of the coating exhibited a monotonic shift toward lower diffraction angles. This systematic leftward shift in peak position serves as direct evidence of macroscopic lattice expansion. Furthermore, Figure 2b presents the Rietveld refinement results for 18YbSZ, with the R-pattern ($R_p$) and R-weighted pattern ($R_{wp}$) values being 3.33% and 5.07%, respectively. Both values are within the acceptable range, confirming the reliability of the refinement results. The crystal structure parameters of 18YbSZ are listed in Table S2, confirming a face-centered cubic (FCC) structure, a structure similar to that of $ZrO_2$, with the space group $Fm\overline{3}m$. This indicates that 18YbSZ is a substitutional solid solution, in which the Zr sites in the lattice are partially substituted by Yb.

When $Yb^{3+}$ ions replace $Zr^{4+}$ ions in the $ZrO_2$ lattice, the system tends to form positively charged oxygen vacancies to maintain macroscopic electrical neutrality[20]. This charge compensation process can be expressed by Equation (1):

$$Yb_2O_3 \xrightarrow{ZrO_2} 2Yb'_{Zr} + V_O^{\cdot\cdot} + 3O_O^{\times} \quad (1)$$

According to Equation (1), for every two $Yb^{3+}$ ions solubilized, one oxygen vacancy is theoretically introduced. To verify this hypothesis, we performed electron paramagnetic resonance (EPR) spectroscopy on plasma-spheroidized YbSZ powder samples. As shown in Figure 2c, all samples exhibited a distinct resonance signal at a g-value of approximately 2.003. EPR signals with g-value in the range of 2.003 are typically attributed to oxygen vacancies[21]. Therefore, it can be confirmed that oxygen vacancies are present in samples with different doping concentrations. Notably, as the $Yb_2O_3$ doping concentration increased, the intensity of the resonance peak at g=2.003 systematically increased, indicating that the concentration of oxygen vacancies increased with the increasing of $Yb_2O_3$ doping concentration.

To determine the chemical valence state of Yb in the YbSZ coating, a detailed spectral analysis of the Yb 4f orbital was performed using X-ray photoelectron spectroscopy (XPS), as shown in Figure 2d. The Yb 4f fine structure was clearly resolved into two peaks: a main peak with a binding energy of 7.37 eV and a satellite peak, which are typical peak positions in the XPS spectra of $Yb^{3+}$ compounds[22,23]. Therefore, it can be confirmed that the Yb in the YbSZ samples exist in the form of $Yb^{3+}$ ions.

Based on the above analytical results, the crystal structure of 18YbSZ was determined, a structural model of the Zr-Yb-O solid solution was constructed, and oxygen vacancies were introduced to maintain charge neutrality, as shown in Figure 2e.

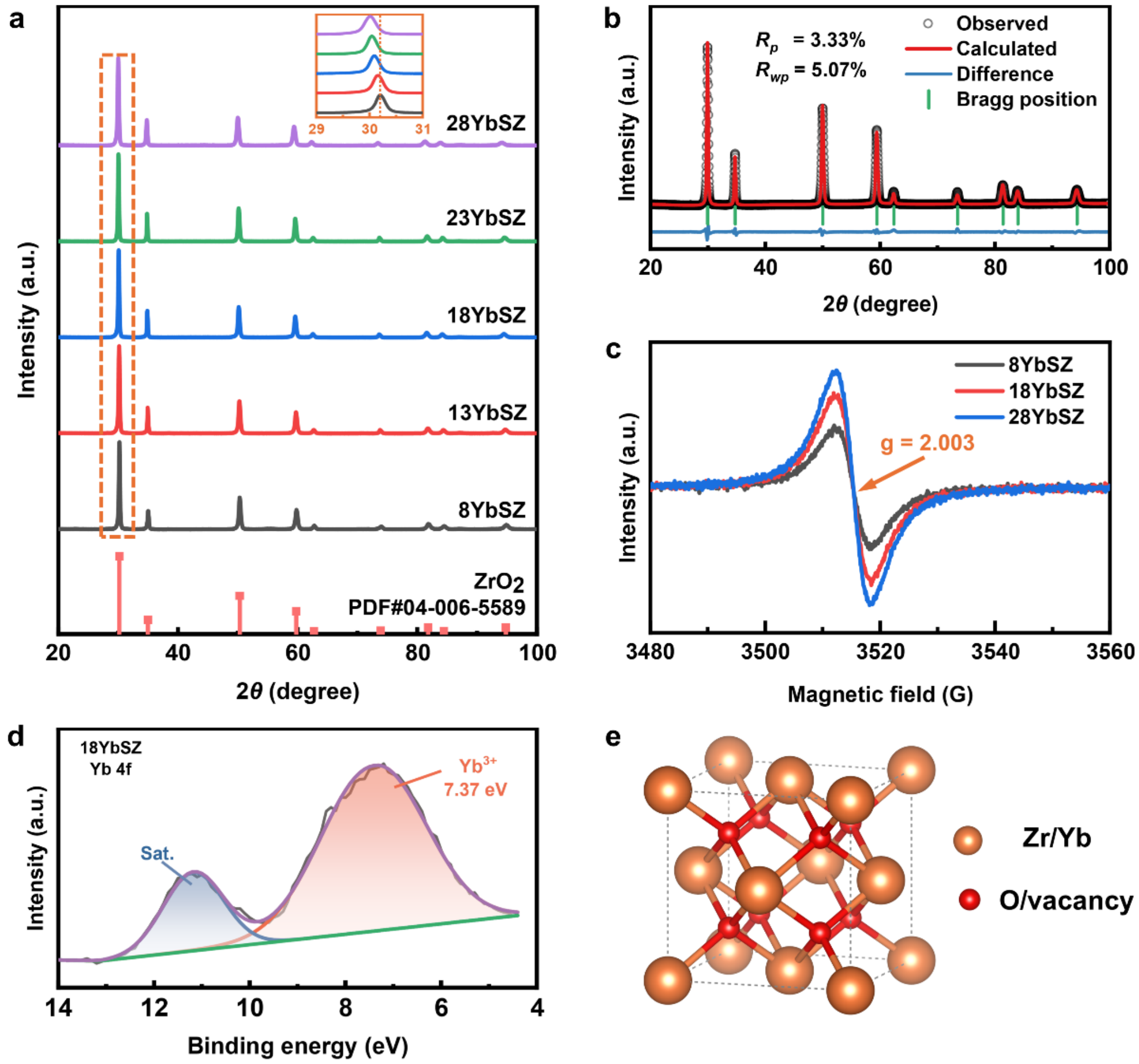

**Figure 2.** Structural and oxygen vacancy defect analysis of YbSZ coatings. a) XRD patterns of YbSZ, confirming the formation of a single-phase solid solution. b) Rietveld refinement of the XRD pattern of 18YbSZ. c) EPR spectra of YbSZ, confirming the presence of oxygen vacancies. d) High-resolution Yb 4f XPS spectrum of 18YbSZ, confirming the +3 valence state of Yb. e) Schematic illustration of the crystal structure of YbSZ.

To further validate the above results for the YbSZ coating, the microstructure of the cross-section was analyzed at the micrometer and nanometer scales, as shown in Figure 3. Figure 3a presents the backscattered electron (BSE) image of the 18YbSZ coating specimen. The cross-sectional image shows that the sample consists of (from bottom to top) a tantalum-tungsten alloy substrate, a Ta10W transition layer, and a YbSZ ceramic top coating. The thickness of the top YbSZ coating is approximately 710 μm. Some pores are observed within the coating, while no distinct vertical through-thickness cracks are evident.

To characterize the crystal structure and the uniformity of elemental distribution at the nanoscale, the 18YbSZ coating was examined using transmission electron microscopy (TEM) and EDS. Figure 3b presents the bright-field (BF) image, revealing multiple distinct grains. EDS mapping reveals a uniform distribution of Zr, Yb, and O, confirming that the solid solution reaction occurred at the nanoscale. Figure 3c shows the selected area electron diffraction (SAED) pattern and high-resolution transmission electron microscopy (HRTEM) image of area A, which has a $[\bar{1}12]$ crystal band axis and exhibits characteristic $(\bar{2}\bar{2}0)$ and $(1\bar{1}1)$ crystal planes (with lattice fringe spacing of 1.83 Å and 2.99 Å, respectively). Figure 3d presents the SAED pattern and HRTEM image of area B. The zone axis of this region is $[130]$. Two sets of lattice fringes corresponding to the $(002)$ and $(\bar{3}11)$ planes are observed, with d-spacings of 2.60 Å and 1.56 Å, respectively. The SAED and HRTEM analyses of two different grains both indicate the formation of a single-phase FCC crystal structure with good crystallinity, providing nanoscale evidence for the formation of a single-phase solid solution.

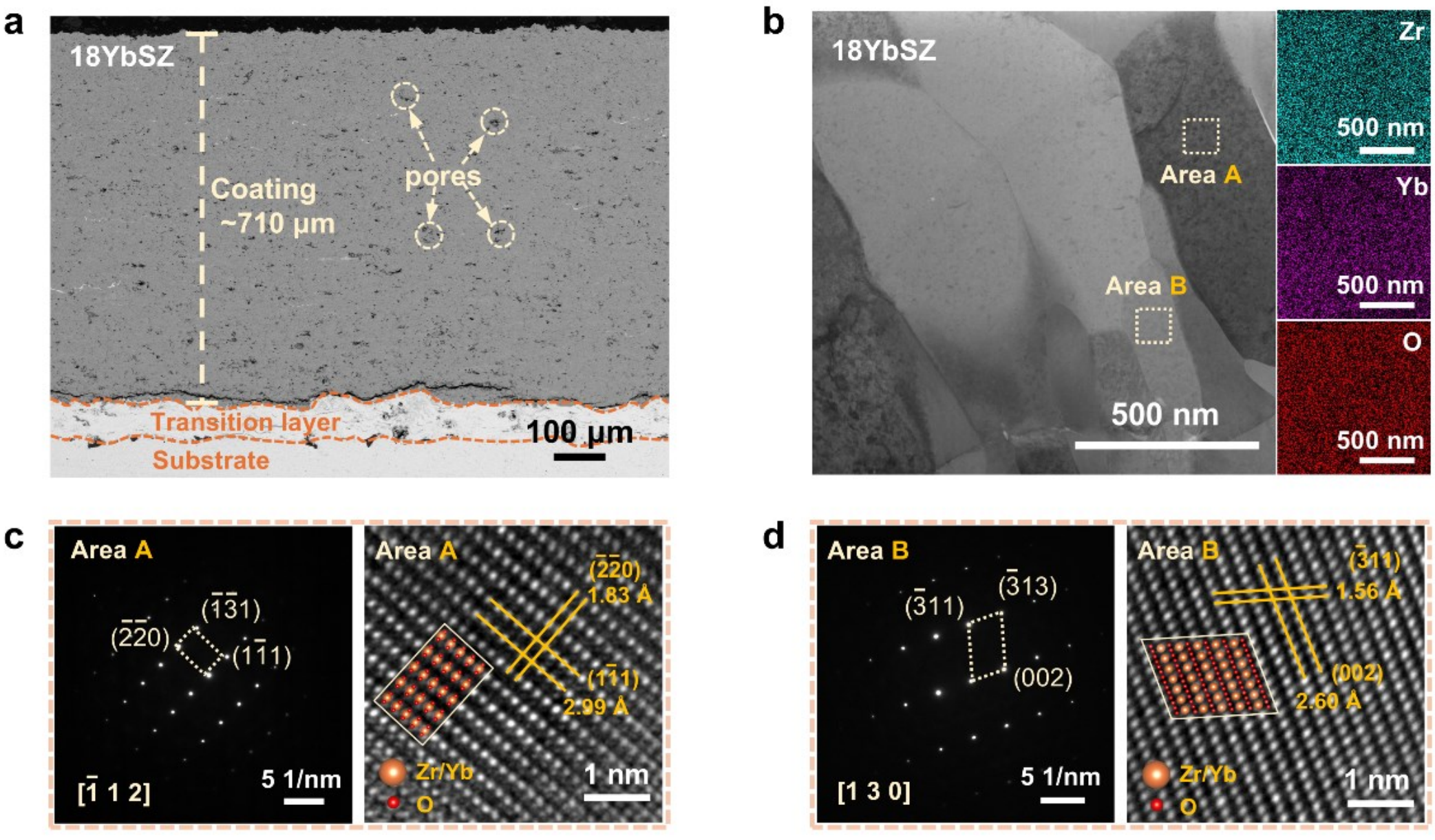

**Figure 3.** Microstructural analysis of the cross-section of the 18YbSZ coating. a) BSE image, showing the microstructure at the micrometer scale. b) BF image and EDS mapping. c) SAED pattern and HRTEM image of area A in (b) (the inset shows the corresponding atomic arrangement). d) SAED pattern and HRTEM image of area B in (b) (the inset shows the corresponding atomic arrangement).

## 2.2 Evaluation of Basic Mechanical and Thermophysical Properties

To preliminarily evaluate the basic properties of the YbSZ coatings, we measured the hardness, elastic modulus, and fracture toughness, and calculated the coefficient of thermal expansion (CTE) via molecular dynamics (MD) simulations. As the Yb content increases, the hardness of YbSZ first rises and then decreases, reaching a maximum at 18YbSZ (Figure 4a). The elastic modulus and hardness exhibit similar trends, with 18YbSZ also showing the highest modulus (Figure 4b). From the standpoint of UHTCs service, a high hardness is highly advantageous[3]. A higher hardness provides greater resistance to plastic deformation and micro-cutting, which significantly reduces the erosion rate and thereby helps preserve the coating thickness and aerodynamic profile[24]. An increase in the elastic modulus implies an enhanced ability to resist elastic deformation, which can reduce the local strain induced by gas pressure and particle impact, thus suppressing the initiation of surrounding cracks[25]. Consequently, at the $Yb_2O_3$ doping content at which hardness and modulus reach their maximum, the coating is expected to exhibit high erosion resistance.

Fracture toughness is a key indicator of a coating's resistance to crack propagation. Figure 4c presents the fracture toughness of the YbSZ coatings. With increasing Yb content, the fracture toughness exhibits an overall decreasing trend, indicating that the solid solution of Yb leads to material embrittlement. When the $Yb_2O_3$ content exceeds 23 mol%, the fracture toughness drops sharply. Under such conditions, even if the coating possesses excellent thermal stability, localized crack propagation may result in overall coating failure[26].

Furthermore, Figure 4d presents the CTE of the YbSZ coatings at various temperatures. Overall, the CTE increases monotonically with increasing Yb content. Due to the limitations of the testing technique, the CTE of the Ta10W alloy substrate was measured only from room temperature to 1600 °C, giving values of $6.6\times10^{-6}$ $K^{-1}$ to $7.3\times10^{-6}$ $K^{-1}$ (Figure S2).Since the CTE of the Ta10W alloy matrix is lower than that of the coating at high temperatures, a higher level of Yb doping leads to a greater thermal expansion mismatch, which often results in localized stress concentration and crack initiation[27]. More unfavorably, the increase in CTE and the monotonic decline in fracture toughness occur concurrently. The synergistic combination of these two detrimental factors severely compromises the service performance of the coating. Hence, only an appropriate amount of $Yb_2O_3$ doping can enhance the thermal protection performance of YbSZ coatings.

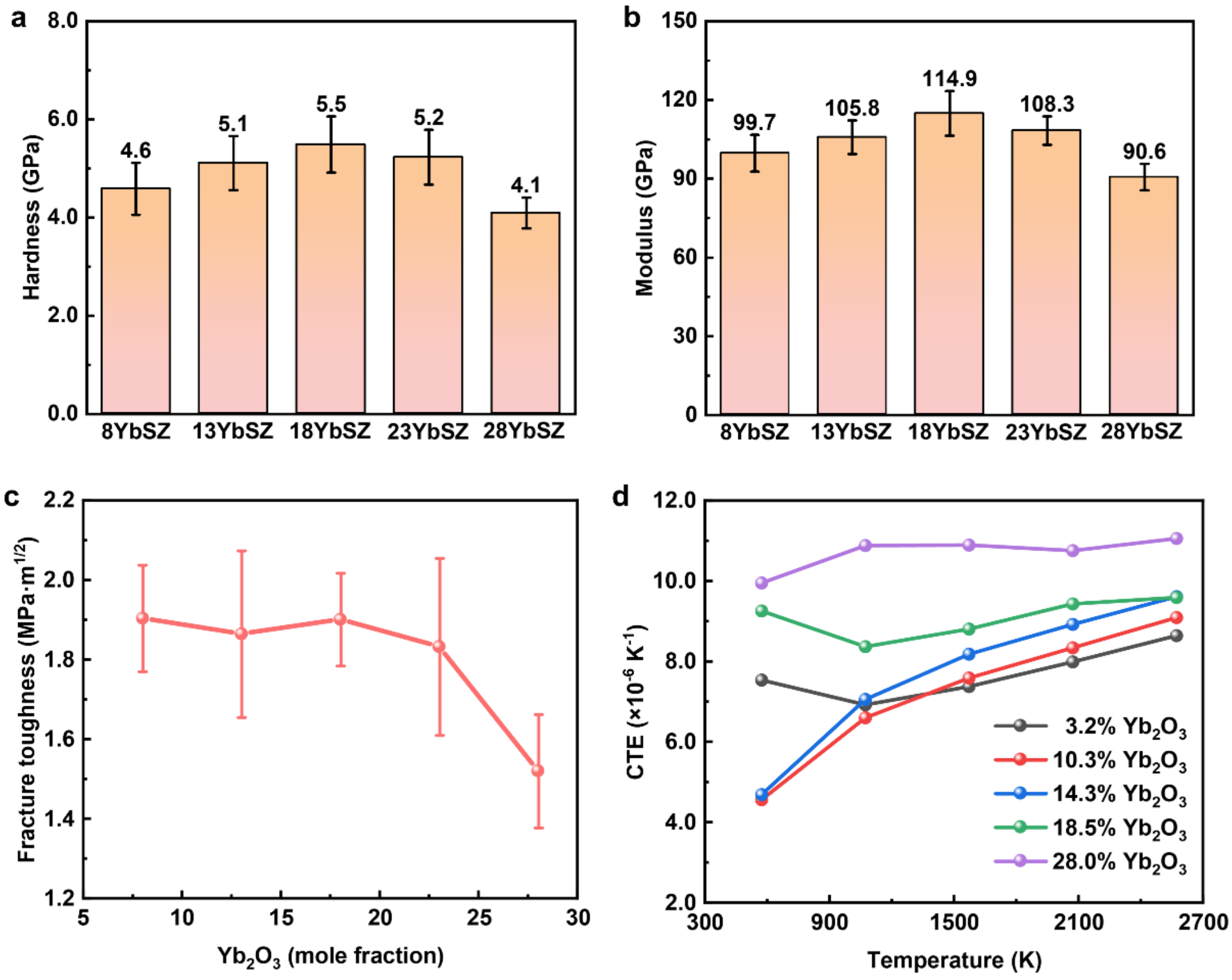

**Figure 4.** Mechanical and physical properties of YbSZ. a) Measured hardness. b) Measured modulus. c) Fracture toughness. d) Calculated coefficient of thermal expansion.

### 2.3 Ablation Behavior and Mechanism under Oxyacetylene and Plasma Conditions

To systematically evaluate the influence of $Yb_2O_3$ doping content on the ultimate ablation resistance temperature of YbSZ coatings and to elucidate the underlying failure mechanisms, oxyacetylene ablation and plasma ablation tests were performed, with the classical thermal barrier coating (5 mol% yttria-stabilized zirconia, YSZ) serving as a reference. Figure 5a shows the temperature profile of the sample surface during the oxy-acetylene ablation process and two typical failure modes. Burn-through failure occurred at 2850 °C for YSZ and at 2990 °C for 13YbSZ, while the 28YbSZ specimen displayed large-area melt recession at 3100 °C. In contrast, no evident failure was observed for the 18YbSZ and 23YbSZ specimens even at the maximum acetylene flow rate. Notably, the maximum surface temperature of both the 18YbSZ and 23YbSZ specimens reached 3182 °C. Under oxyacetylene ablation conditions, the 18YbSZ specimen withstood ablation for 100 s at surface temperatures exceeding 3100 °C without any evident failure, demonstrating strongest ultra-high temperature thermal protection performance compared with all the reported results.

The temperature of the sample surface during the plasma ablation process is shown in Figure 5b. During the tests, the current and gas flow rates were progressively increased until melt blow-off occurred on the coating surface. Under plasma ablation conditions, the conventional YSZ coating exhibited a maximum temperature of 2716 °C. With a certain amount of $Yb_2O_3$ doping, the 8YbSZ coating showed a slight increase in its upper temperature limit to 2752 °C. As the $Yb_2O_3$ content was further increased, the upper temperature limits of the 13YbSZ, 18YbSZ, and 23YbSZ coatings all approached ~2800 °C, demonstrating markedly superior ablation resistance. However, when the $Yb_2O_3$ content became excessive, the upper temperature limit of the 28YbSZ coating dropped sharply to 2644 °C, which was even lower than that of the YSZ coating. These results indicate that an appropriate amount of $Yb_2O_3$ doping can effectively elevate the ablation-resistance temperature limit of YbSZ coatings, whereas excessive doping leads to a drastic deterioration in their thermal protection performance.

The results above indicate that samples with different $Yb_2O_3$ doping levels exhibit significantly different ablation resistance under both ablation environments. Specifically, 18YbSZ can withstand thermal loads of nearly 3200 °C during oxy-acetylene ablation without suffering complete failure; however, under plasma ablation conditions, the coating fails due to melting at temperatures approaching 2780 °C.

Oxy-acetylene ablation provides a relatively stable, macroscopically uniform high-temperature, strongly oxidizing environment[28]. As a stable oxide, the YbSZ sample itself undergoes almost no oxidation in a strongly oxidizing atmosphere; therefore, failure is not primarily caused by oxidative erosion. However, the low thermal conductivity of the oxide coating induces a steep internal temperature gradient, and the resulting thermal mismatch stress promotes crack initiation and propagation[29]. Concurrently, inherent pores in the coating provide diffusion pathways for oxygen, causing the transition layer or substrate to gradually oxidize, ultimately leading to interfacial failure, which manifests macroscopically as localized holes. When the surface temperature increases further and substantially exceeds the melting point, the coating undergoes melting, softening, and dripping.

In contrast, plasma ablation represents a more severe ablation condition, characterized by an extremely high heat flux, intense transient thermal shock, and concurrent mechanical scouring[30]. Under the experimental conditions, the heat flux of oxyacetylene ablation was approximately 2.4 $MW/m^2$, while that of plasma ablation reached 11 $MW/m^2$. Furthermore, according to previous studies[31,32], the flow velocities of the oxyacetylene and plasma flames are approximately 200 m/s and as high as 450 m/s, respectively. This significant difference in both heat flux and flow velocity directly influences the dominant failure mechanisms. Under plasma ablation conditions, the coating surface is rapidly heated to a temperature exceeding its melting point. The high-velocity gas flows then scours the molten layer, causing melt blow-off, which makes surface melting the dominant failure mode. The upper temperature limit measured in plasma ablation is therefore closely correlated with the melting point of the material.

The ablated 18YbSZ sample was further characterized by TEM, confirming that the Zr-Yb-O solid solution structure did not decompose and demonstrating the compositional uniformity of individual elements at the nanoscale. Figure 5c shows a BF image of the 18YbSZ specimen after ablation, captured near a grain boundary. Figure 5d presents the SAED pattern, the BF image, and the corresponding atomic model of area C. Indexing revealed that the zone axis is $[001]$, and the lattice fringe spacings of the $(200)$ and $(0\bar{2}0)$ planes are 2.56 Å and 2.58 Å, respectively. The measured interplanar spacings along the two orthogonal directions differ by only 0.02 Å, corresponding to a relative deviation of less than 0.8%. This slight discrepancy is attributed to the local lattice distortion induced by the solid solution of Yb atoms and the presence of oxygen vacancies. Figure 5e presents the nanoscale elemental distribution at the grain boundary shown in Figure 5c, confirming the compositional uniformity of individual elements without apparent enrichment. Based on these results, it can be concluded that the ablated sample retains a single-phase FCC crystal structure, with no evidence of phase transformation or decomposition relative to the pre-ablation sample. This demonstrates the stabilizing effect of Yb on $ZrO_2$ and confirms that the failure of the YbSZ coating during plasma ablation is caused by the congruent melting of the solid solution.

To further demonstrate its advantages, the present study compares the results of 18YbSZ with those from a series of similar investigations (Figure 5f), highlighting its remarkable enhancement in the upper ablation-resistance temperature limit. It should be noted that although the total ablation duration for the YbSZ coating was 300 s, the temperature was increased progressively during the test; therefore, the entire 300 s duration cannot be directly used for comparison. Here, we define the performance of 18YbSZ as withstanding ablation at temperatures exceeding 3100 °C for 100 s without evident failure. The ultimate ablation temperatures of most material systems reported in the literature are below 2600 °C, and those exceeding 3000 °C are even rarer. A comparison of this work with UHTCs having an ablation-resistant temperature above 2600 °C reveals fundamentally different performance enhancement mechanisms. Existing strategies for improving coating performance have primarily focused on two aspects: optimization of the coating/substrate interfacial structure and regulation of oxidation-product phase evolution. The former enhances interfacial bonding and alleviates thermal stress through the construction of graded transition layers, thereby suppressing interfacial delamination[28]. The latter improves the protective capability of oxide scales by tailoring the flow behavior of molten glassy phases to promote defect sealing and self-healing, for example, through the self-healing effect of molten $SiO_2$ or by introducing $TiO_2$ to reduce melt viscosity and facilitate crack filling[33,34]. By comparison, the enhancement mechanism in this work directly targets the essence of ablation—the melting point of the oxidation product—thereby circumventing structural degradation caused by the loss of low-melting-point phases or active oxidative evaporation. This “melting point elevation” strategy provides a new paradigm for overcoming the thermal protection challenges under ultra-high temperatures accompanied by high-speed gas scouring.

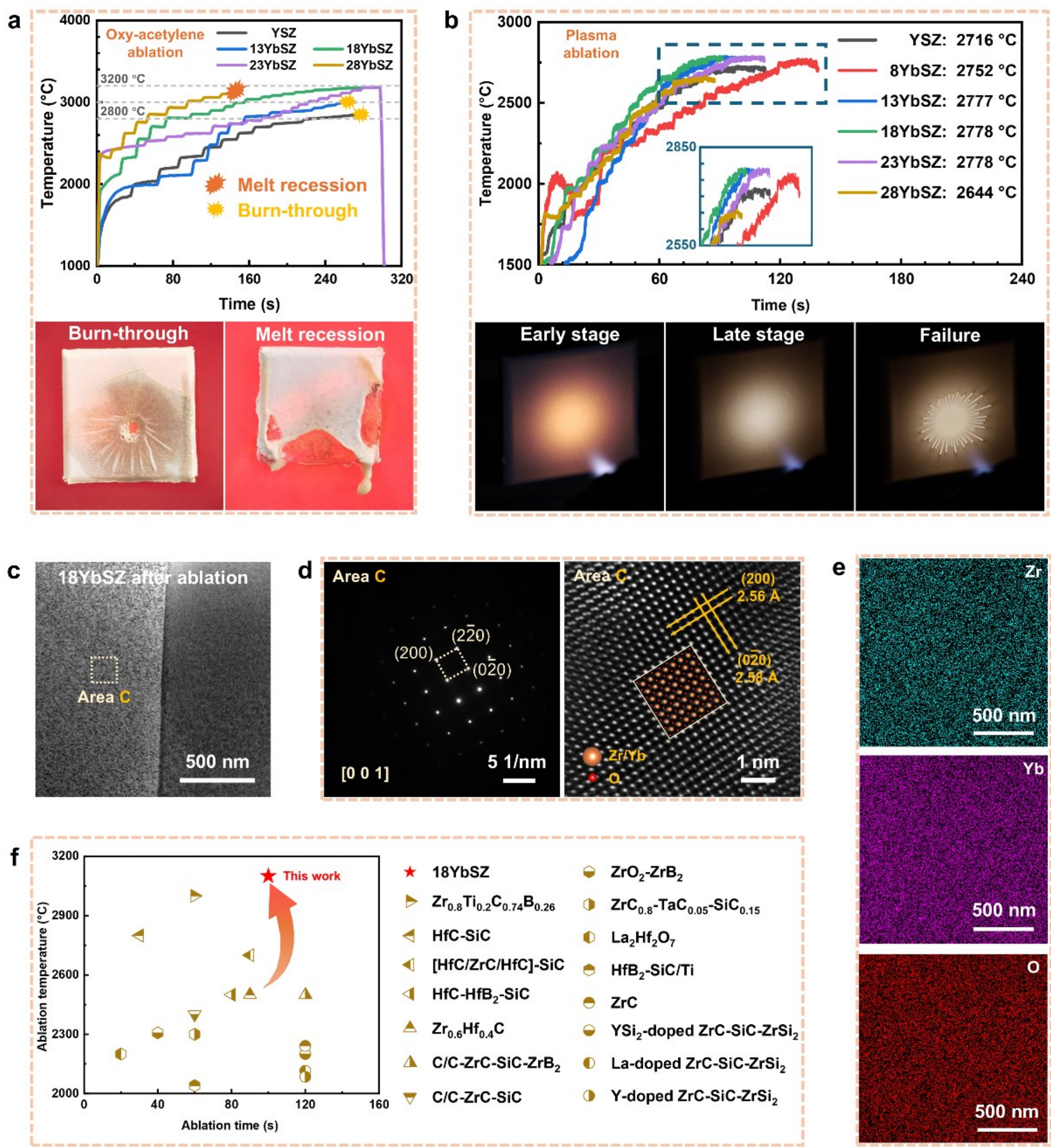

**Figure 5.** Ablation performance and microstructural characterization of the ablated region for YbSZ. a) Temperature profiles of oxyacetylene ablation and macroscopic images of failure. b) Temperature profiles of plasma ablation and macroscopic images captured during the ablation process (early stage, late stage, and failure). c) BF image of the ablated region, showing a grain boundary between two grains. d) SAED pattern and HRTEM image of area C in (c) (the inset shows the corresponding atomic arrangement). e) EDS mapping of the region in (c), showing the nanoscale compositional uniformity of individual elements. f) Comparison between this work and a series of existing similar studies, demonstrating the superiority in ablation performance[28,33–46].

## 2.4 Mechanism of Melting Point Elevation

To elucidate the mechanism underlying the ablation resistance, the melting points of YbSZ samples were measured (Figure 6a). It can be seen that YbSZ exhibits an “anomalous solid solution” effect: unlike other oxides that depress the melting point of $ZrO_2$, $Yb_2O_3$ elevates it. Specifically, the melting point of YbSZ increases non-monotonically with the increase of $Yb_2O_3$ content, reaching a maximum of ~2850 ℃ within 15-20 mol% range. These trends correlate strongly with the ablation behavior of YbSZ, confirming the crucial role of the high melting point in enabling superior high-temperature ablation resistance. To unravel this anomalous solid solution behavior, the defect chemistry of YbSZ with 6.7, 10.3, 14.3, 18.5, and 23.1 mol% $Yb_2O_3$ were investigated by first-principles calculations. As shown in Figure 6b, the oxygen vacancy formation energy in YbSZ is initially negative and decreases with increasing $Yb_2O_3$ content, reaching a minimum of -3.615 eV at 14.3 mol%, indicative of the most thermodynamically favorable defect formation. Beyond this composition, the formation energy becomes positive and rises sharply, implying markedly reduced thermodynamic stability for vacancy generation. These results suggest that, at low doping levels, sparsely distributed oxygen vacancies stabilize the cubic fluorite structure of YbSZ and thereby elevate its melting point. However, once the $Yb_2O_3$ content exceeds a critical threshold, excessive oxygen vacancies and substituted Yb ions destabilize the lattice, leading to a subsequent decline in melting temperature.

It is well known that higher ionic charges enhance electrostatic interactions, thereby stabilizing the crystal lattice[47]. As shown in Figure 6c, the net charge on Yb ions in YbSZ exhibits a non-monotonic dependence on $Yb_2O_3$ content, peaking at 1.378 e at 14.3 mol%. This trend closely parallels the variation of the oxygen vacancy formation energy, indicating that enhanced Yb-O bonding directly elevates the stability of YbSZ. Figure 6d presents the charge density isosurface on the (111) plane of YbSZ doped with 14.3 mol% $Yb_2O_3$, where the color coding at atomic centers denotes elemental types only. The overall electron distribution exhibits limited charge overlap between cations and anions, confirming the predominantly ionic bonding character. Notably, enhanced charge density is evident in the Yb-O regions (highlighted by the yellow dashed line) compared to the Zr-O regions, indicating stronger covalent interactions associated with Yb substitution. These results further indicate the critical role of $Yb_2O_3$ concentration in modulating the structural stability of YbSZ, and thus its melt point.

Figure 6e further presents the partial density of states (PDOS) of YbSZ at the optimal doping level (14.3 mol% $Yb_2O_3$). The valence band is predominantly composed of O 2p states, while the conduction band primarily derives from Zr 4d and Yb 5d orbitals, reflecting the characteristic electronic structure of an ionic compound[48]. Notably, significant hybridization between Yb 5d and O 2p states is observed in the energy range of -5 eV to -2 eV. This orbital overlap reveals substantial covalent character in Yb-O bonding, which reinforces local chemical interactions. The synergistic coupling of ionic and covalent bonding markedly enhances interatomic cohesive forces, thereby improving the lattice's resistance to thermal shock. Consequently, $Yb_2O_3$ incorporation exerts a dual regulatory effect on lattice stability by tuning both the local electronic structure and defect chemistry. Moderate doping concurrently strengthens ionic and covalent interactions, stabilizes oxygen vacancies, and elevates thermal stability. Conversely, excessive $Yb_2O_3$ induces bond weakening and defect destabilization, ultimately compromising lattice robustness. This mechanistic insight provides a robust theoretical foundation for optimizing the thermal protection performance of YbSZ coatings.

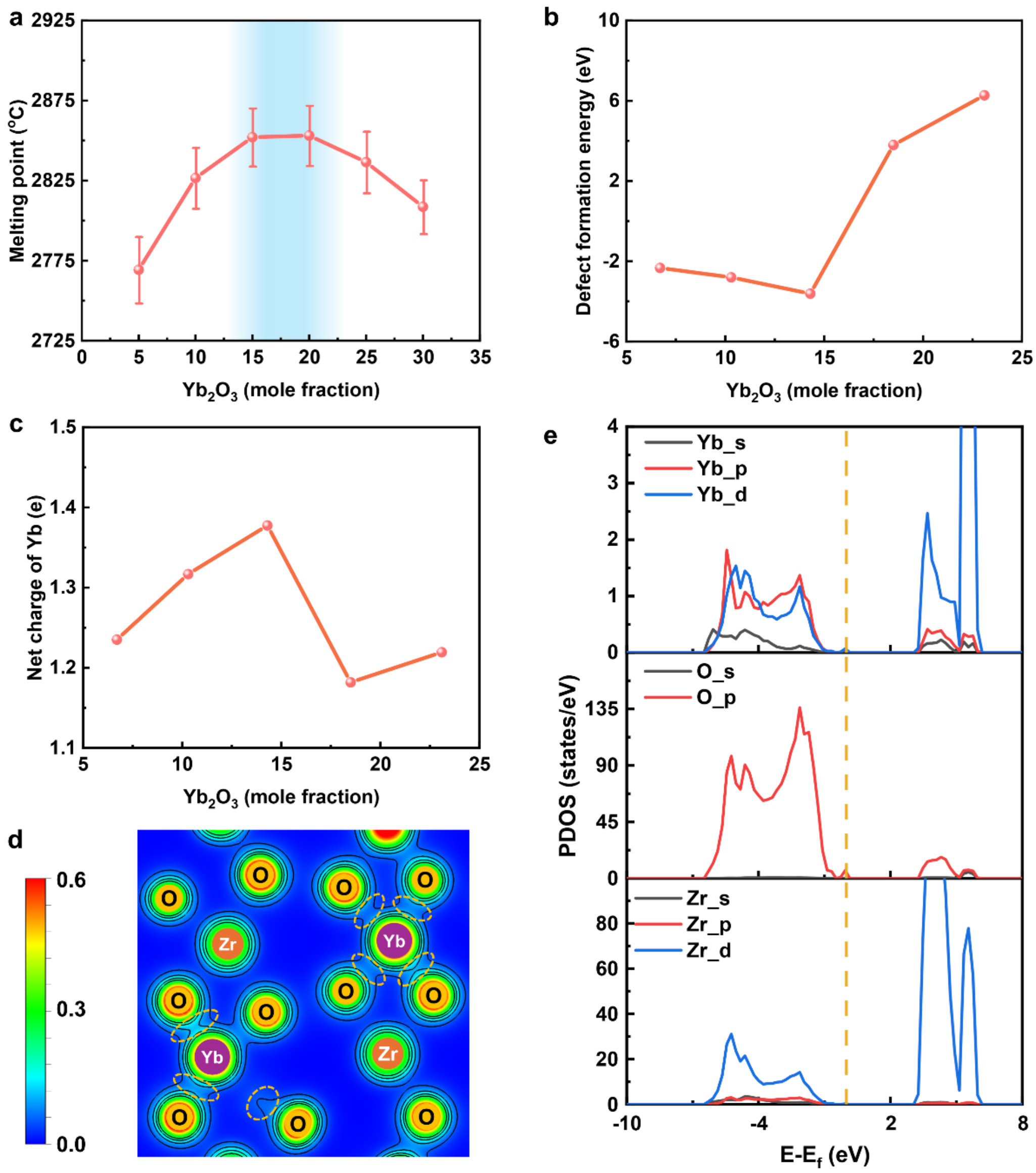

**Figure 6.** Melting point measurements and the mechanism of anomalous solid solution in YbSZ. a) Melting point, b) defective formation energy, and c) net charge of Yb in YbSZ. d) Charge distribution on the (111) crystal plane and e) partial orbital density (PDOS) of YbSZ doped with 14.3 mol% $Yb_2O_3$.

## 3 Conclusion

In summary, we have successfully developed a novel $Yb_2O_3$ stabilized $ZrO_2$ solid solution coating materials, realizing the "anomalous solid solution" phenomenon in oxide ceramics. This achievement raises the melting point of $ZrO_2$-based ceramics to approximately 2850 °C and extends the ultimate plasma and oxyacetylene ablation temperature up to nearly 2780 °C and 3180 °C, respectively. This enhancement results from the addition of an appropriate amount of Yb to the solid solution, which synergistically enhances both the ionic and covalent bonds while also increasing the stability of oxygen vacancy defects. More importantly, this study establishes a universal design paradigm for ultra-high temperature ceramic materials. The proposed concept of "anomalous solid solution" can be extended to other UHTC systems, enabling the development of boride, carbide, and carbonitride ceramics through inverse design. Future research should focus on the toughening of oxide ceramics and the exploration of new ceramic systems based on "anomalous solid solution" anions, in order to address the cracking issue in oxide ceramics.

# 4 Method

## 4.1 Material Synthesis

High-purity $ZrO_2$ and $Yb_2O_3$ powders (purity: 99.9%, particle size: ~500 nm, Beijing Huawaruike Chemical Co., Ltd., China) were added to heated deionized water at a solid loading of 50 wt.%. Subsequently, 1 wt.% polyethylene glycol (PEG, Shanghai Aladdin Biochemical Technology Co., Ltd., China) was introduced to obtain a mixed slurry. The slurry was milled using a ball mill mixer at 450 rpm for 10 h. Near-spherical $ZrO_2$-$Yb_2O_3$ micro-sized composite powders were then produced by a spray granulation apparatus at an atomizer frequency of 50 Hz. The collected powders were subsequently heat-treated in a muffle furnace at 1400 °C for 4 h, followed by atmospheric plasma spheroidization; the key parameters are listed in Table S3. Finally, the resulting composite powders were sieved to obtain the desired particle size fraction suitable for plasma spraying.

Ta10W alloy (Beijing Zhongke Yannuo New Materials Technology Co., Ltd., China) plates measuring 30×30×5 mm were used as substrates. Prior to spraying, the substrate surfaces were degreased with anhydrous ethanol and subsequently grit-blasted with corundum to enhance surface roughness. To further increase surface roughness and improve bonding strength, a Ta10W transition layer was deposited between the substrate and the YbSZ coating. Both the YbSZ coating and the bond coat were applied via atmospheric plasma spraying, and the corresponding key parameters are listed in Table S4. The reference sample used 5YSZ ($ZrO_2$–5 mol% $Y_2O_3$, particle size: 15~45 μm, Beijing Huawaruike Chemical Co., Ltd., China).

### 4.2 Characterization and Basic Properties Testing

The flowability and apparent density of the powders were measured using a Hall flowmeter. The funnel had a cone angle of 60° and an orifice diameter of 2.5 mm. For the flowability test, 50.0g of the test powder was accurately weighed, and the time required for the entire powder to flow through the funnel orifice was recorded. The measurement was repeated three times, and the arithmetic mean was calculated. If the powder failed to flow freely, it was classified as "non-flowable". The apparent density was determined using the same apparatus, additionally equipped with a cylindrical measuring cup of 25.0 $cm^3$ capacity. The center of the funnel orifice was aligned with the center of the cup, and the distance between the bottom of the funnel and the top of the cup was set to 25 mm. The powder was allowed to flow freely into the cup until it was completely filled; the excess powder was then scraped off, and the mass of the powder inside the cup was weighed. The apparent density ($\rho$) was calculated using the following equation:

$$\rho = \frac{m}{V} \tag{2}$$

where $m$ is the mass of the powder in the measuring cup (g), and $V$ is the volume of the cup ($cm^3$). Each sample was measured three times, and the arithmetic mean was taken.

The phase structure of the samples was analyzed by X-ray XRD (D8 ADVANCE, Bruker) with Cu Kα radiation. The XRD patterns were refined using GSAS-II software[49] and the Rietveld method[50]. Oxygen vacancies were analyzed by EPR (A300, Bruker) on plasma-spheroidized 8YbSZ, 18YbSZ, and 28YbSZ powders. The measurements were conducted at a cryogenic temperature of 4 K provided by liquid argon. The valence state of Yb was analyzed by XPS (K-alpha, Thermo Scientific). A fine-scan of the Yb 4f region was performed on the 18YbSZ coating, and the spectra were calibrated and fitted using the instrument-integrated Avantage software. Owing to the severe spectral overlap between the conventional strong peaks of Zr 3d and Yb 4d, characterizing the Yb 4f region is an effective method for confirming the valence state. The unit cell structures were visualized using the VESTA software[51]. The microstructure and elemental distribution were characterized by SEM (MIRA LMS, TESCAN) and EDS (Ultim Max, Oxford Instruments), respectively. TEM lamellae of the 18YbSZ samples before and after ablation were prepared using a FIB (Scios 2 DualBeam, Thermo Scientific). TEM (FEI Talos F200) was employed to characterize the nanostructure, and the HRTEM images were further analyzed using DigitalMicrograph.

The elastic modulus and hardness of the materials were measured using a nanoindenter (G200X, KLA) in quasi-static loading mode. For each sample, six test points were randomly selected to obtain load-displacement curves. The Young's modulus and hardness were derived from the experimental data based on the classic Oliver-Pharr (O-P) model[52]. Subsequently, the fracture toughness was calculated by an energy-based method using the load-displacement curve data[53,54]. The CTE of the Ta10W substrate was measured using a push-rod dilatometer (DIL 402C, NETZSCH, Germany).

The melting point test is conducted using the recalescence phenomenon[55]. Because the material releases latent heat during solidification, a fluctuation or plateau appears on the cooling curve, and this thermal event was used to identify the melting point. The melting point measurements were carried out using a 350 W sealed $CO_2$ laser heater with a wavelength of 10.6 μm. The material was first heated until it melted, after which the heating was stopped. An infrared pyrometer (Pyroskop 840, Kleiber) with a response time of 10 μs was then used to record the temperature during the subsequent cooling and solidification, from which the cooling curves were obtained. The emissivity was calibrated using the emissivities of $ZrO_2$ and $Yb_2O_3$; the melting points of pure $ZrO_2$ and $Yb_2O_3$ were determined to be 2710 °C and 2413 °C, respectively[56,57]. The emissivity of the YbSZ samples was approximately calculated using a mixing rule.

**4.3 Ablation Testing**

In the oxyacetylene ablation test, the distance from the flame gun nozzle to the initial specimen surface was 10 mm. During the test, the surface temperature response of the specimen was recorded using an infrared pyrometer (MR1SCSF, Raytek). The initial flow rates of acetylene and oxygen were set to 10.8 L/min and 14.4 L/min, respectively. The acetylene flow rate was then progressively increased until reaching a maximum of 19.2 L/min, at which point the oxygen flow rate was 16.8 L/min. The test was terminated if pronounced melting or burn-through on the coating surface. If no evident failure occurred at the maximum acetylene flow rate, the test was stopped when the ablation duration reached 300 s.

In the plasma ablation test, the distance from the nozzle to the initial specimen surface was 60 mm. The surface temperature response of the specimen was recorded using a dual-band infrared pyrometer (METIS M322, Sensortherm). The plasma flame was generated by exciting a mixture of argon and nitrogen gases, with argon and nitrogen flow rates of 60 L/min and 25 L/min, respectively. The ablation temperature was controlled by modulating the output power of the plasma system through current regulation. The maximum power applied to the specimens prior to melting ranged from 67.8 kW to 76.5 kW. Notably, the maximum power applied to the 18YbSZ specimen during ablation was 76.5 kW, with a current of 818.7 A and a voltage of 93.5 V.

**4.4 DFT calculation and MD simulation**

All DFT calculations in this work were performed using the Vienna Ab initio Simulation Package[58,59]. The electron-ion interactions were described by the projector augmented wave (PAW) method[60]. The exchange-correlation effects were treated using the strongly constrained and appropriately normed (SCAN) meta-generalized gradient approximation (meta-GGA) functional[61]. The plane-wave cutoff energy was set to 520 eV. The Brillouin zone was sampled using a Γ-centered scheme with a k-point spacing of approximately 0.3 $Å^{-1}$, except for the DOS, where it was refined to 0.1 $Å^{-1}$. On the basis of the $Fm\bar{3}m$ $ZrO_2$ unit cell, 2 × 2 × 3 supercells were generated using the special quasi-random structure (SQS) method implemented in the Alloy Theoretic Automated Toolkit (ATAT), to ensure the random distribution of Yb substitution sites and oxygen vacancies while maintaining the charge neutrality of the system[62]. All supercells were fully relaxed, and the energy and the force convergence criteria were set to $10^{-6}$ eV and 0.01 eV/Å, respectively.

The defective formation energy ($E^{f}_{[V_O]}$) for oxygen vacancies is defined by Equation (3)[63]:

$$E^{f}_{[V_O]} = E^{tot}_{defect} - E^{tot}_{bulk} + n\mu_O \tag{3}$$

where $E^{tot}_{defect}$ and $E^{tot}_{bulk}$ denote the total energies of the supercells with and without oxygen vacancies, respectively. The $n$ and $\mu_O$ are the number of oxygen vacancies and the energy of an oxygen atom in the oxygen molecule, respectively.

Based on the optimized YbSZ crystal structure, a high-precision static self-consistent calculation was performed to obtain an accurate valence electron charge density. Subsequently, the Bader charges ($Q_{\text{Bader}}$) of the Yb atom were calculated[64]. The net charge of a Yb atom ($\Delta Q_{\text{Yb}}$) is defined by equation (4), where $Z_{\text{val}}$ is the nominal number of valence electrons for Yb in the pseudopotential used.

$$\Delta Q_{\text{Yb}} = Z_{\text{val}} - Q_{\text{Bader}} \tag{4}$$

MD simulations were performed using GPUMD with the NEP89 general neural evolution potential to investigate the coefficient of thermal expansion of the material[65,66]. The 4 × 4 × 4 simulation systems were constructed based on the supercells established by the SQS method. The system was equilibrated in the NPT ensemble at each temperature for 1 ns using a stochastic cell rescaling barostat with a time step of 1 fs. The equilibrium lattice constant was taken as the average lattice parameter over the last 10 ps. Since YbSZ adopts a cubic fluorite structure and exhibits isotropic macroscopic thermal deformation, the linear thermal expansion coefficient (α) relative to 300 K can be defined according to Equation (5):

$$\alpha = \frac{1}{a_{300\,K}} \frac{a_T - a_{300\,K}}{T - 300} \tag{5}$$

Where $a_{300\,K}$ is the equilibrium lattice constant at 300 K, and $a_T$ is the equilibrium lattice constant at temperature $T$.

## Acknowledgements

The authors gratefully acknowledge the Institute of Carbon Matrix Composites, Henan Academy of Sciences, for providing the oxyacetylene ablation testing apparatus.

# Supplementary Materials

**Table S1.** Apparent density and flowability of 18YbSZ powders prepared by different processes

| Process | Apparent density ($g/cm^3$) | Flowability (s/50g) |
|---|---|---|
| Spray granulation | 1.08 | 109.3 |
| After heat treatment | 1.36 | non-flowable |
| After atmospheric plasma spheroidization | 2.93 | 43.3 |

**Table S2.** Crystal structure parameters of 18YbSZ from Rietveld refinement of XRD

| $Zr_{0.64}Yb_{0.36}O_{1.9742}$ | Space group | Unit cell | | | Main characteristic peaks | | |
|---|---|---|---|---|---|---|---|
| | | a (Å) b(Å) c(Å) | α(°) β(°) γ (°) | Volume ($Å^3$) | h k l | d (Å) | Position (°) |
| Crystal structure parameters | *F m -3 m* | 5.145664 5.145664 5.145664 | 90 90 90 | 136.246 | 1 1 1 | 2.9709 | 29.894 |
| | | | | | 2 0 0 | 2.5728 | 34.681 |
| | | | | | 2 2 0 | 1.8193 | 49.937 |
| | | | | | 3 1 1 | 1.5515 | 59.373 |
| | | | | | 2 2 2 | 1.4854 | 62.309 |
| | | | | | 4 0 0 | 1.2864 | 73.402 |
| | | | | | 3 3 1 | 1.1805 | 81.298 |
| | | | | | 4 2 0 | 1.1506 | 83.887 |
| | | | | | 4 2 2 | 1.0504 | 94.172 |

**Table S3.** Key Processing Parameters in Powder Preparation

| Process | Related parameters | | | | |
|---|---|---|---|---|---|
| Spray granulation | Inlet temp./°C | Outlet temp./°C | Frequency/Hz | Rotational speed/rpm | |
| | 200 | 120 | 50 | 10 | |
| Heat treatment | Temperature/°C | Time/h | | | |
| | 1400 | 4 | | | |
| Atmospheric plasma spheroidization | Primary gas (L/min) | Auxiliary gas (L/min) | Voltage /V | Electric current/A | Powder feed rate (g/min) |
| | 65～70 | 20～25 | 93~98 | 700～750 | 18～20 |

**Table S4.** Key Processing Parameters in Coating Preparation

| Process | Primary gas (L/min) | Auxiliary gas (L/min) | Voltage /V | Electric current/A | Powder feed rate (g/min) |
|---|---|---|---|---|---|
| Ta10W transition layer | 55～60 | 20～25 | 90~95 | 550～600 | 12～15 |
| YbSZ coating | 55～60 | 20～25 | 90~95 | 630~650 | 10~12 |

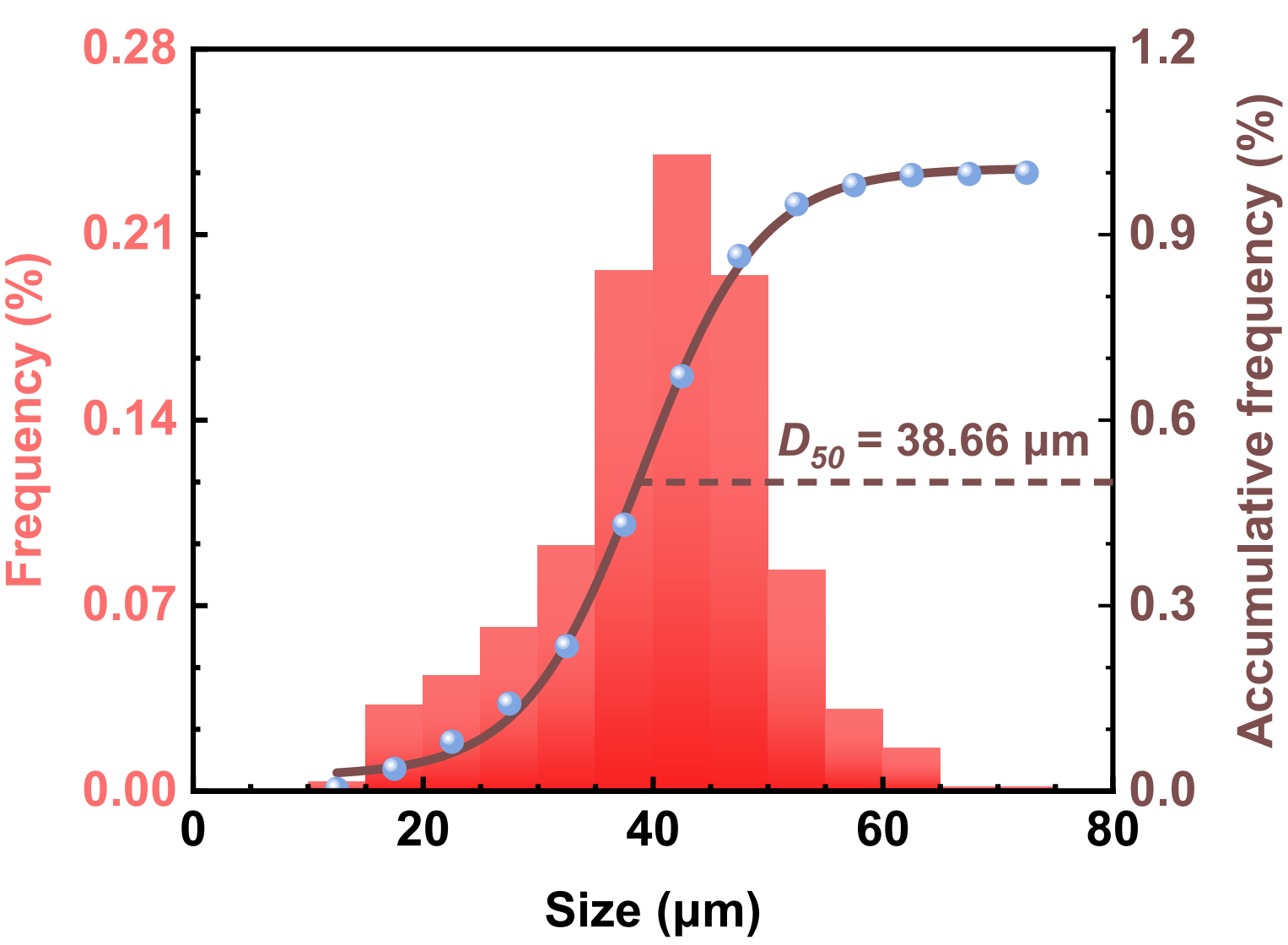

**Figure S1.** Particle size distribution of 18YbSZ powder ultimately used for plasma spraying.

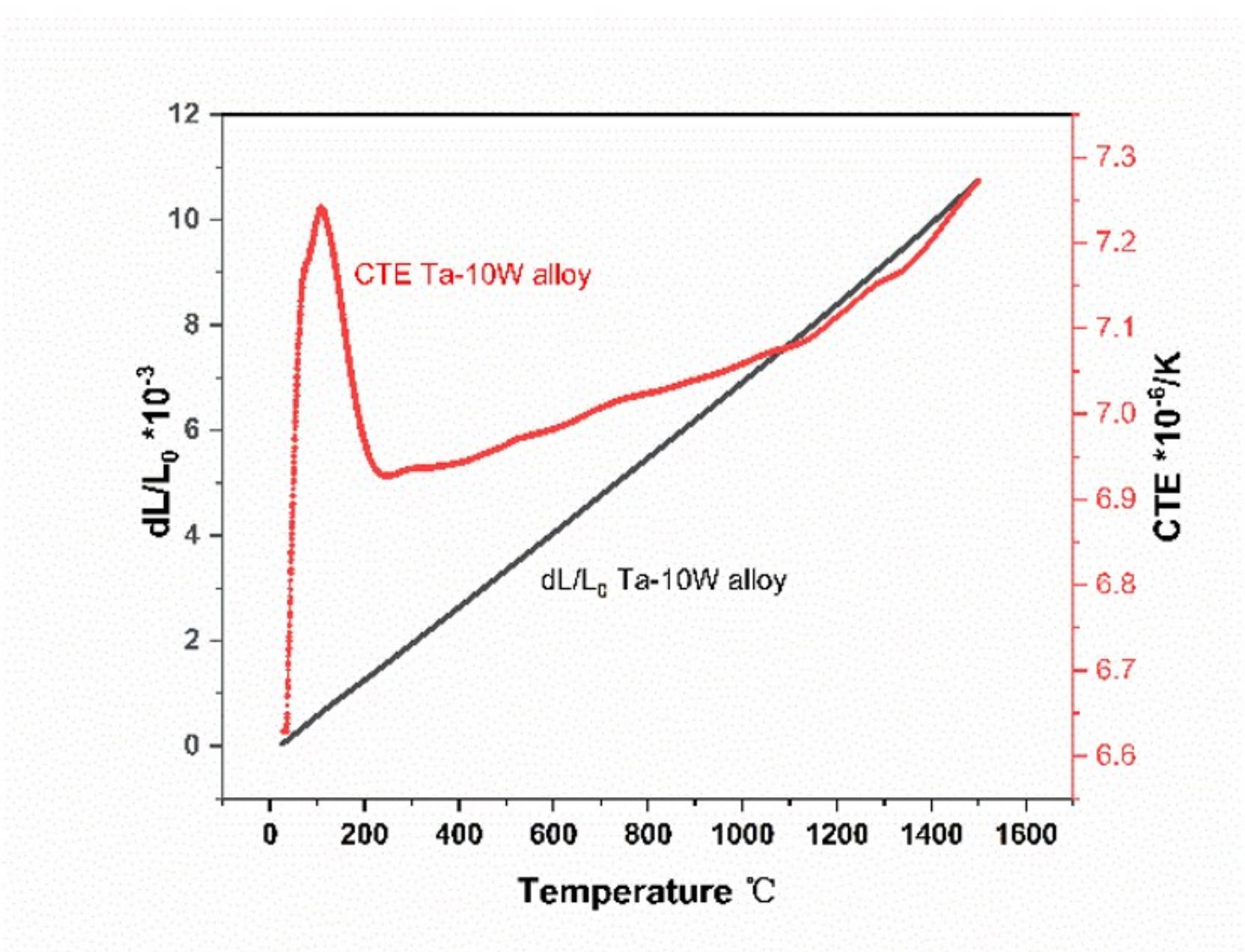

**Figure S2.** The coefficient of thermal expansion (CTE) of the Ta10W alloy from room temperature to 1600 ℃.